\documentclass[runningheads]{llncs}
\usepackage[T1]{fontenc}
\usepackage[numbers,sort&compress,square]{natbib}

\usepackage{xurl}
\usepackage{booktabs}
\usepackage{array}
\usepackage{ragged2e}

\usepackage{graphicx}
\usepackage[hidelinks]{hyperref}

\begin{document}
\title{IKTRACE: A Natural Language Processing Framework for Identifying Indigenous Knowledge Representation in Academic Literature}

%
\author{
Tara Azin\inst{1}\thanks{Corresponding author: \email{tazin@gcrc.carleton.ca}} \and
Christy Caudill\inst{1} \and
Mako Sorensen\inst{1} \and
Peter Pulsifer\inst{1}
}

\authorrunning{T. Azin et al.}

\institute{
Geomatics and Cartographic Research Centre (GCRC), Carleton University \\ Ottawa, Canada
}
\titlerunning{IKTRACE}
%
%
\maketitle              
\begin{abstract}

As artificial intelligence systems increasingly rely on vast textual data, respectful and accurate representation of Indigenous Knowledge has become critical. This paper introduces IKTRACE, a natural language processing framework for identifying, analyzing, and organizing Indigenous Knowledge representations in academic literature while adhering to Indigenous Data Sovereignty (IDsov) principles. We present a community-informed taxonomy co-developed with Inuit partners, comprising nine knowledge domains grounded in Indigenous Data Sovereignty principles. Building on this taxonomy, we implement a multi-method Knowledge identification pipeline using complementary natural language processing approaches. Outputs are integrated into a knowledge graph to support future expert and community validation. Applied to a corpus of Canadian Arctic research documents, IKTRACE demonstrated high accuracy and strong community relevance in expert validation. Our framework provides a scalable, ethically grounded foundation for Community-centered Knowledge validation and the development of Indigenous-informed AI systems.

\end{abstract}

\section{Introduction}

Indigenous Knowledge systems represent thousands of years of ecological observation, cultural practices, and governance structures that have sustained communities and ecosystems across generations. In recent decades, academic researchers have increasingly recognized the value of Traditional Ecological Knowledge (TEK) and Indigenous Knowledge (IK) for addressing contemporary challenges in conservation, climate change adaptation, and sustainable resource management~\cite{maracle2025conservation}. This growing body of research contains valuable information that, when properly identified and structured, could support accurate representation of IK on the web and in AI systems. However, as AI-based technologies increasingly harvest web content to train Large Language Models (LLMs), construct knowledge graphs, and populate crowdsourced platforms like Wikidata, the need for approaches to identify, verify, and assess this information has become critical~\cite{duncan2024benchmark,wiechetek2024ethical}. There is an urgent need to examine publicly accessible academic resources to identify IK representations that respect Indigenous Data Sovereignty (IDsov) principles~\cite{kukutai2016sovereignty, rda2019care,carroll2020care2}, enable community validation, and can be appropriately incorporated into digital knowledge systems. Such examination ensures that information from ethically conducted research can be made accessible in ways that adhere to IDsov principles, including consent, attribution, and community control, rather than perpetuating colonial patterns of knowledge extraction~\cite{carroll2020care2}.

This paper presents a novel approach to IK identification and representation analysis through three key contributions. First, we introduce a co-developed community-informed taxonomy comprising 9 domains, developed through partnerships with Indigenous organizations and aligned with OCAP\textsuperscript{\textregistered} (Ownership, Control, Access, and Possession) principles\footnote{OCAP\textsuperscript{\textregistered} principles assert First Nations' control over data collection and ownership of how their information is used~\cite{fnigc_ocap}.}. This taxonomy serves both as a conceptual framework for understanding IK representation and as a practical foundation for respectful information extraction frameworks.

Second, we develop a multi-method information extraction pipeline combining three complementary NLP approaches, Named Entity Recognition (NER), semantic similarity search, and language model-based structured extraction, followed by keyword extraction from the outputs. The extracted keywords are integrated into a knowledge graph that supports data organization, community validation, and future accessibility of identified IK representations.

Third, we implement a validation framework that incorporates expert review of extracted representations to verify appropriateness and alignment with Indigenous priorities for Knowledge access and sovereignty. All models used in this pipeline were closed-source models accessed via secure APIs, and all data analyzed consisted of publicly available literature. Our approach maintains strict adherence to community data sovereignty principles by ensuring no unauthorized access to protected or sensitive Indigenous Knowledge.

\section{Background and Motivation}
\subsection{Data Sovereignty and CARE}

Indigenous data sovereignty asserts the inherent right of Indigenous communities to govern the collection, access, and use of data concerning their lands, cultures, and knowledge systems~\cite{spano2025sovereignty}. This concept challenges ``data colonialism'' or its more recent form, ``digital colonialism'', where actors use Indigenous information, such as Traditional Knowledge, for profit without community control or benefit \cite{oneil2024cld,spano2025sovereignty}. In response, the Global Indigenous Data Alliance (GIDA) developed the CARE principles (Collective Benefit, Authority to Control, Responsibility, and Ethics) in 2019 as a framework for Indigenous data governance that addresses significant gaps in existing global data standards~\cite{gida_global}. In the age of rapid AI advancement, adhering to CARE is vital for protecting intangible cultural heritage from misappropriation and ensuring that computational approaches support IK sovereignty rather than generating inaccurate or harmful representations~\cite{moshagen2024ilt}.

The CARE principles support Indigenous Data Sovereignty by prioritizing people and purpose-centric AI development over purely data-driven approaches~\cite{junker2024goldrush}. The framework's four pillars provide practical mechanisms for protecting Indigenous interests in technology development. Real-world implementations show these principles in action. Te Hiku Media, a Māori tribal radio station, developed the ``Kaitiakitanga license'' to govern use of their language data, rejecting corporate offers to purchase recordings and ensuring community control~\cite{spano2025sovereignty}. Similarly, the Algonquian Dictionaries Project~\cite{junker_algonquian} enforces strict policies against unauthorized data mining, and in one case they demanded deletion of improperly extracted data from AI repositories~\cite{junker2024goldrush}. These examples illustrate how CARE principles provide practical mechanisms for protecting Indigenous Knowledge sovereignty in the age of AI.

\subsection{Cultural Knowledge Preservation in the Age of AI}

Preserving Indigenous Cultural Knowledge presents critical challenges as current AI systems prioritize efficiency and standardization over cultural authenticity. This results in significant loss of contextual meaning and nuance~\cite{anik_translation,tian2022recontextualization}. The problem is compounded by a data gap where Indigenous Knowledge remains underrepresented in digital spaces, often existing in limited forms created by non-community members that fail to represent authentic cultural perspectives~\cite{wiechetek2024ethical}. AI systems trained on Western paradigms strengthen knowledge hierarchies that exclude Indigenous epistemologies~\cite{duncan2024benchmark}, while digital colonialism enables corporate actors to appropriate Indigenous Knowledge without permission. AI development driven by standardized benchmarks prioritizes technical metrics over community needs, and recent language models pose barriers through homogenization, creating feedback loops where AI-generated content lacking cultural depth distorts knowledge transmission~\cite{mihalcea2025weird}. These challenges expose serious gaps in research practices that fail to prioritize Indigenous Data Sovereignty and Community-defined preservation goals.

\subsection{Gaps in Current Practice}

Critical gaps remain in current AI and machine learning practices concerning Indigenous Cultural Knowledge and representation. A fundamental issue is the assumption that large data volumes will eventually encode all necessary cultural knowledge for AI training. This data-hungry architecture is incompatible with IK systems, which exist in limited digital forms and cannot be adequately represented through scale alone~\cite{moshagen2024ilt}. At the same time, the lack of ethical oversight in data sourcing creates tension between the AI community's demand for more data and Indigenous Communities' rights to data sovereignty.

Meanwhile, ethically produced resources developed with community consent and expert validation do exist, but they are often overlooked because identifying and verifying them requires deeper understanding of IK systems. These materials must also be examined to understand how Indigenous Communities are positioned in discourse. To address these challenges, this work proposes a taxonomy of key IK elements to support the identification, validation, and appropriate incorporation of existing resources into AI systems.

\section{Co-developing an Indigenous Knowledge Taxonomy}

Co-development methodology requires that the research agenda encompasses a diverse array of approaches to guarantee equitable participation and engagement from all participants (e.g., rights holders and stakeholders)~\cite{norstrom2020coproduction,yua2022coproduction}. Indigenous communities are not merely stakeholders or interest groups, but are rights holders with decision-making authority, as has been affirmed by domestic (Sections 25, 35, 91, and 92 of the Canadian Constitution Act~\cite{canada_constitution}) and international frameworks (United Nations Declaration on the Rights of Indigenous Peoples; UNDRIP~\cite{un_un_drip}). A strong foundation of trust with rights holders in partnered communities, with their organizations and people, is a long-term effort of respectful relations. Our research group and Pitquhirnikkut Ilihautiniq (the Kitikmeot Heritage Society, KHS) have collaborated for over a decade on projects including those to meet their digital strategy priorities. Building on this long-term relationship, KHS and the Iqaluktuuttiaq community partnered with our research group and the Arctic Institute of North America (AINA) to carry out shared objectives under the Arctic Genomics Project~\cite{arcticgenomics}. The Arctic Genomics project brought together diverse partners across disciplines, cultures, and organizations with a focus on biodiversity monitoring, wildlife genome information that might support Inuit-led monitoring, and co-management of wildlife that are key to the social, cultural, physical, and economic well-being of Inuit in the Canadian Arctic and in Inuit Nunangat, largely. Once the research partnerships were formed between the rights holders and other relevant actors, a Social Science and Traditional Knowledge Research License from the Nunavut Research Institute was granted. Ethics approval for this study was obtained from the Carleton University Research Ethics Board.

\begin{table}[htbp]
\centering
\caption{Indigenous Knowledge Taxonomy Domains with example terms. The complete taxonomy developed through co-development sessions contains a more extensive list of terms. Representative examples are shown here due to space constraints.}
\label{tab:taxonomy}
\small
\setlength{\tabcolsep}{6pt}
\renewcommand{\arraystretch}{1.25}
\begin{tabular}{
>{\RaggedRight}p{0.28\linewidth}
>{\RaggedRight}p{0.34\linewidth}
>{\RaggedRight\arraybackslash}p{0.34\linewidth}
}
\toprule
\textbf{Domain} & \textbf{Description} & \textbf{Example Terms} \\
\midrule

\textbf{Knowledge Holders and Communities} &
Indigenous peoples, knowledge holders, and representative organizations &
Elders; knowledge holders; Inuit Circumpolar Council; Nunavut Research Institute \\
\addlinespace[0.5em]

\textbf{Governance and Sovereignty} &
Governance systems, rights, legal frameworks, and power relations &
Self-governance; Inuit laws; food sovereignty; right to self-determination \\
\addlinespace[0.5em]

\textbf{Indigenous Knowledge Systems} &
Traditional knowledge, cultural practices, spirituality, and language &
Traditional ecological knowledge; cultural revitalization; Inuktitut; spirituality \\
\addlinespace[0.5em]

\textbf{Land, Environment, and Resources} &
Territories, environmental relationships, and land-based activities &
Inuit Nunangat; socio-ecological systems; harvesting; wildlife management \\
\addlinespace[0.5em]

\textbf{Food Systems and Sovereignty} &
Food security, traditional foods, and food practices &
Country food; food sovereignty; traditional diet; safe butchering practices \\
\addlinespace[0.5em]

\textbf{Health and Well-Being} &
Holistic health, traditional medicine, and health monitoring &
Traditional medicines; well-being; disease monitoring; contaminants \\
\addlinespace[0.5em]

\textbf{Cultural Materials and Practices} &
Traditional clothing, tools, and material cultural expressions &
Ulu handle; kayak; drumming; feasts \\
\addlinespace[0.5em]

\textbf{Research and Collaboration} &
Research processes, collaborative practices, and ethical frameworks &
Co-management; meaningful engagement; genomics monitoring; community-based reporting \\
\addlinespace[0.5em]

\textbf{Socio-Cultural Concepts} &
Social relations, values, and contemporary challenges &
Respect; climate change; seasonality; continual planning and preparation \\
\bottomrule
\end{tabular}
\end{table}

As part of the Arctic Genomics effort, Inuit partners expressed that the preservation of Inuinnaqtun, an endangered Inuktut dialect, and proper ownership, control, and attribution (e.g., crediting an Elder for the development of a new Inuinnaqtun term) of the language and terms were primary priorities. Inuit partners asserted that Indigenous Data Sovereignty includes knowledge and language as well as the right to control how they are positioned when their knowledge and language are written and shared.

The primary engagement setting for the development of the taxonomy for this study was during a community-based and ecosystem-level genomics workshop (MAGIC: Muskoxen and Genomics in the Community) held at the Canadian High Arctic Research Station (CHARS), Cambridge Bay, NU, from January 9--12, 2024 under the Arctic Genomics project~\cite{mishra2025magic}. During the workshop, Inuit participants detailed how muskoxen are essential to the culture, economies, and livelihoods of northern Indigenous communities across Inuit Nunangat, insisting that research be grounded in Inuit ways of knowing and carried out to enhance and broaden Inuit sovereignty.  Muskoxen is a culturally important species for the Iqaluktuuttiaq community and the Kitikmeot Region, and served as a concept through which Inuit and non-Indigenous researchers could communicate about socio-ecological complexities embedded in Inuit culture and ways of knowing. From these knowledge sharing sessions, a socio-ecological taxonomy was developed. Online Zoom and phone discussions took place between academic researchers and community participants, developing the taxonomy in an iterative approach between March 2024 and December 2025. The terms and structure of the taxonomy are provided in Table~\ref{tab:taxonomy}.

The process employed in this study to co-develop an Indigenous Knowledge taxonomy differs radically from the development of a colonial worldview taxonomy. FAIR and Linked Open Data (LOD) taxonomies are commonly developed through committee process and available for open use (e.g., Environmental Ontology (ENVO) provides precise and controlled vocabularies of environments and related processes~\cite{buttigieg2016envo,envo_browse}). During our co-development process,  Inuit partners defined terms in their own reference frame and knowledge system, rather than only relying on a pre-conceived list of terms relevant in western knowledge domains, attending to ontological Indigenous self-determination. In addition to CARE data principles, non-Indigenous researchers in this study were careful to attend to protocols including OCAP\textsuperscript{\textregistered}~\cite{fnigc_ocap}, ITK National Inuit Strategy on Research~\cite{itk_nisr}, and the ICC Circumpolar Inuit Protocols for Equitable and Ethical Engagement~\cite{icc2022protocols}, all of which predate and inform the creation of the CARE Principles~\cite{gida_care_nationtalk}.

\section{Methodology for IK Representation Analysis}

\begin{figure}[t!]
\centering
\includegraphics[width=0.8\textwidth]{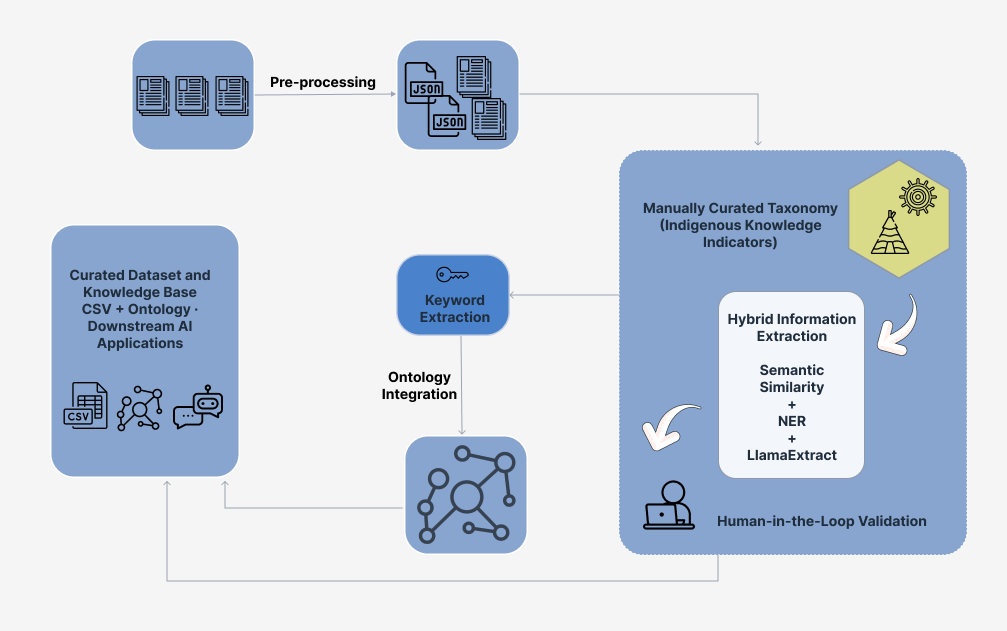}
\caption{Pipeline architecture for identifying Indigenous Knowledge representations in academic literature. Documents are preprocessed, then analyzed using three NLP methods (NER, semantic similarity search, and LLM-based analysis). Keywords from all outputs are integrated into a knowledge graph for expert and community validation.}
\label{fig2}
\end{figure}

This section presents the methodology we used to analyze how Indigenous Knowledge is represented in academic literature and how we identified, partially validated, and stored these data for final validation and use in AI systems in the future. Building upon the IK taxonomy, our approach combines three complementary NLP methods to identify entity-level references, discourse patterns, and structured knowledge representations across our corpus. The output was then used to extract keywords that were then integrated into a knowledge graph. This graph provides the foundation for a future chatbot system informed by TEK and IK instances that were collected from academic literature and validated by experts and Indigenous community partners in Canada. Our approach prioritized both re-usability of identified and extracted information and the importance of organized annotation and validation to support future analysis and community access.  Figure~\ref{fig2} illustrates the complete pipeline architecture. The following subsections describe each component in detail.

\subsection{Dataset Construction}

We compiled a corpus of 417 documents including academic articles and reports related to IK and TEK in the Canadian Arctic. The documents were sourced from two primary initiatives: the Arctic Genomics project~\cite{arcticgenomics}, which supports conservation and natural resource management of Arctic wildlife, and the Northern Contaminants Program (NCP)~\cite{northerncontaminants}, established in 1991 to address human exposure to contaminants in traditional food sources. All documents were publicly available through open-access journals or government repositories. The corpus totalled approximately 31,000 pages, with NCP documents substantially longer than typical journal articles as they often compile multiple research studies within single documents.

The corpus included topics such as Arctic genomics, Arctic ecosystems, wildlife genomics and conservation, environmental health, and responses to Northern contaminants. Given the geographic focus on Northern communities and their traditional territories, these documents contain extensive references to Indigenous communities. The corpus includes both peer-reviewed journal articles and government-funded research reports in Canada, published between 1990 and 2023.

To ensure optimal performance of our NLP pipeline, we implemented several preprocessing steps. We converted all PDF documents to JSON format, with each document segmented into paragraphs while preserving metadata including the source filename, page numbers, and document structure. We removed non-textual elements including figures, tables, and reference lists, as these sections could introduce noise into the extraction process. We also excluded preliminary sections (tables of contents, acknowledgments) and bibliographic sections to focus on main body text.\footnote{Datasets and implementation details are publicly available at \url{https://github.com/IndigenousKnowledge/IKTRACE}.}

\subsection{Information Extraction Pipeline}

\subsubsection{Named Entity Recognition (NER)}
Named entity recognition is a core NLP task that identifies and classifies named entities in text into predefined categories. For our purpose, this approach allowed us to identify entities such as Indigenous Communities, Knowledge Holders, organizations, and territories referenced in the literature. By quantifying entity mentions and extracting their surrounding context, this method enabled analysis of how frequently these entities were referenced and the discourse patterns in which they appeared.

In this work, we used \texttt{GLiNER-medium}, a zero-shot NER model that identifies arbitrary entity types without fine-tuning~\cite{zaratiana2024gliner}. We defined 20 entity labels organized into four categories: Knowledge Holders, organizations, geographic regions, and Knowledge domains.\footnote{The complete taxonomy is available in the project repository: \url{https://github.com/IndigenousKnowledge/IKTRACE/tree/main/data/taxonomy}.} Text pre-processing included citation removal and a minimum content threshold of 45 words. The model was configured with a confidence threshold of 0.4. For each extracted entity, we recorded the entity text, label, confidence score, and a three-sentence context window to preserve semantic meaning. \texttt{GLiNER} extracted more than 100K entities across all documents, then stored them in structured format with source metadata for validation and ontology integration. An expert with extensive domain knowledge and close working relationships with Indigenous Community partners reviewed 20\% of the annotated items, and the acceptance rate for the identified entities was 94.87\%.

\subsubsection{Semantic Similarity Search}
To identify IK representations beyond named entities, we performed semantic similarity search using sentence-level embeddings. This method enables detection of discourse patterns, which we used to examine how Indigenous Communities are positioned within the literature. We used the outputs to explore whether communities tended to be framed as active contributors whose knowledge informed research, or as passive subjects affected by external factors. By identifying these patterns, we gain insight into how Indigenous agency, knowledge systems, and sovereignty are represented in the literature, which may inform both the present analysis and future community-driven evaluations of research practices.

We used the \texttt{all-MiniLM-L6-v2} sentence transformer~\cite{wang2020minilm}, a lightweight model optimized for semantic similarity tasks. We defined 11 semantic categories aligned with our taxonomy, each represented by a set of seed terms. The categories explore various dimensions of Indigenous community representation including identity recognition, agency framing (active versus passive positioning), Knowledge valorization (strength-based versus deficit-based representations), and the contextualization of Indigenous rights, governance systems, and environmental relationships.

For each category, we computed a prototype embedding by averaging the embeddings of its seed terms and normalizing the resulting vector. Sentences were extracted from processed paragraphs using spaCy~\cite{honnibal2020spacy}. Semantic similarity scores were computed using cosine similarity between sentence embeddings and category prototypes. We applied a similarity threshold of 0.5 based on pilot tuning to balance precision and recall trade-offs while preserving semantically relevant IK signals. Sentences exceeding this threshold were retained with a three-sentence context window (preceding, target, and following sentences). The semantic similarity search extracted over 6,500 matches across all categories \footnote{Text snippets may match multiple category prototypes and receive multiple labels. The reported count represents total matches, not unique snippets. This overlap applies across all components of the extraction pipeline.}.

To validate the matches, three experts who work closely with Indigenous communities (two university professors and one PhD student) reviewed 35 items (105 items in total). They assessed match accuracy, suggested re-categorizations where needed, evaluated community usefulness, and provided additional insights. We elaborate on these findings in section \ref{Discussion}.





\subsubsection{LlamaExtract} We used LlamaExtract, a data extraction service from LlamaIndex~\cite{liu2022llamaindex}, to complement the other discourse-level extractions. LlamaExtract is an extraction service that uses large language models with schema-based prompting to extract predefined information fields from documents.

We defined a custom extraction schema consisting of 10 fields aligned with our taxonomy: Indigenous group, Indigenous organization, conservation location, Indigenous person, conservation strategy, Indigenous language, traditional territory, Inuit law, socio-ecology, and traditional knowledge. Each field included explicit descriptions to guide the extraction model.The extraction agent was configured via the Llama Cloud API in FAST mode with per-page extraction.\footnote{The underlying hosted base model is abstracted by the service and is not user-specified. See \url{https://www.llamaindex.ai}.} System prompts emphasized extracting exact text without paraphrasing to preserve original phrasing and cultural context. Using this tool, over 64,000 instances were identified across the corpus. This approach covered specific knowledge types that might be missed by NER or semantic similarity methods, particularly complex multi-word expressions and domain-specific terminology that were embedded within longer passages.

\subsubsection{Keyword Extraction}

To complement the terms extracted in the NER step, we applied keyword extraction to text snippets from the semantic similarity and LlamaExtract outputs. We used KeyBERT~\cite{grootendorst2020keybert}, a keyword extraction technique that leverages deep learning embeddings to identify keywords by measuring their semantic similarity to each text snippet. More than 105K terms were extracted across all documents. These extracted terms provide the basis for the ontology integration step described in the following subsection.

\subsection{Ontology Integration}
Integrating identified IK representation terms into a structured ontology, or knowledge model, has multiple critical purposes. First, it provides structured organization of diverse knowledge types, facilitates validation and review by community experts, supports discovery of relationships between concepts, and ensures long-term preservation and accessibility of identified knowledge. Second, the ontology provides a framework for connecting Indigenous concepts with existing Western academic discourse while maintaining the semantic richness and contextual relationships that are essential for culturally appropriate representation.

\begin{figure}[t!]
\centering
\includegraphics[width=0.75\textwidth]{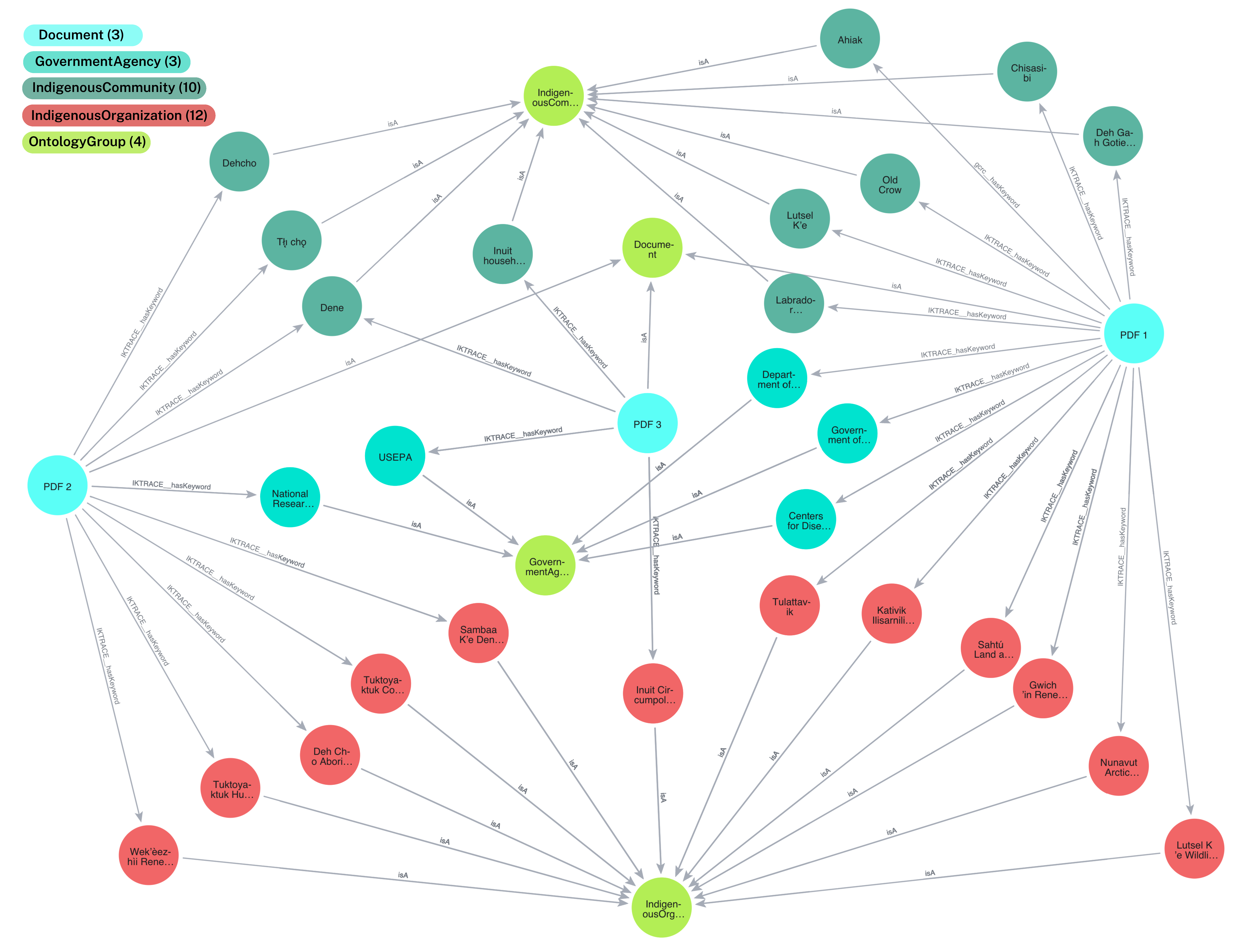}
\caption{Example knowledge graph linking three document instances to ontological concepts via \texttt{isA} relations.}
\label{fig1}
\end{figure}

Our ontology integration consists of two components, the ontology and the knowledge graph. The ontology represents the general structure that groups different concepts together, whereas the knowledge graph contains the instances that populate the ontology. In practice, an academic article is represented in the ontology and has a specific instance (e.g., PDF1.pdf) in the knowledge graph. This supports adherence to interoperability standards and enables connection to the semantic web and RDF frameworks, while providing flexibility to represent complex relationships and inform users effectively. The platform is built on Node.js and Express, with Neo4j serving as the graph database and modeler. Objects are stored on our local server, and Node.js handles connections to Neo4j through standard queries. Figure~\ref{fig1} represents an example of the knowledge graph structure for three sample documents where instances are connected to the ontology through \texttt{isA} relations.

The knowledge graph prioritizes relationships between concepts rather than isolated data points. This supports nuanced links between Indigenous Knowledge and Western academic contexts where one-to-one correspondences may not exist.  It also enables efficient navigation of concept relationships and provides relational data that can support language models without extensive training data.

\section{Discussion and Conclusion}
\label{Discussion}

We presented a framework for identifying and analyzing Indigenous Knowledge representations in academic literature through a community-informed taxonomy using an NLP pipeline. This work responds to priorities articulated by Inuit partners, who emphasized that Indigenous Data Sovereignty includes knowledge and language as well as the right to control how communities are positioned when their knowledge is written and shared. Our approach prioritizes respect for Indigenous Data Sovereignty principles and protection as its primary commitment. By eliciting insights and feedback from experts who maintain close relationships with community members, we address critical gaps in identifying, categorizing, and storing Indigenous Knowledge using AI-based tools and approaches. This structured process has two main purposes. First, it prepares information for direct validation by community members themselves. Second, it supports development of AI tools informed by Indigenous Knowledge systems for community use, and ensures that Indigenous Knowledge in existing AI systems is accurately represented based on knowledge frameworks validated by the community.

\subsection{Validation and Community Relevance}

At this stage, validation was conducted through expert-mediated review by researchers working closely with Indigenous communities, with direct community-led validation planned as the next phase of the project. Expert validation of the outputs of our pipeline demonstrates both high accuracy and community relevance. The 94.87\% NER acceptance rate demonstrates that zero-shot models configured with culturally informed categories can effectively identify community-related entities and cultural references in AI systems handling sensitive cultural data.

For semantic similarity search, we conducted a more comprehensive validation with three experts working closely with Indigenous communities. We sampled 105 items (at least three samples per category) from the semantic similarity outputs and divided them into three sets of 35 items per evaluator. Experts evaluated whether assigned categories were meaningful, suggested alternative categorizations when appropriate, and assessed potential community usefulness. Results showed that 82 items (78\%) were deemed useful for community access, with 68 items (65\%) correctly categorized. This validation process helped us verify label comprehensibility, gather expert judgment from community-engaged researchers, assess community utility, and identify distribution considerations for non-Indigenous audiences.

Expert assessors raised important concerns that inform future development. First, they emphasized the need to explicitly recognize the extent of the contribution of the Indigenous community in the source documents. Second, they recommended consistent capitalization of terms such as ``Knowledge'' to show respect. Third, they noted the importance of carefully handling terminology such as ``Aboriginal'', ``Indigenous'', and specific community names to prevent misunderstanding and support culturally appropriate usage. 

\subsection{Limitations and Future Work}

Our approach addresses the challenge of identifying ethically produced Indigenous Knowledge in academic literature, collecting it for community assessment, and establishing a foundation for community use and development of IK-informed AI systems with reliable, accurate information. By focusing on publicly available collaborative research and prioritizing identification of knowledge representations rather than extraction, we enable communities to assess their representation and control digital accessibility. However, several limitations exist. First, our corpus focuses on Canadian Arctic literature with a taxonomy co-developed with Inuit partners that may not represent other Indigenous nations' knowledge systems. Second, reliance on existing literature intentionally excludes sacred or undocumented knowledge to respect sovereignty. Finally, pipeline performance depends on source document quality, requiring ongoing critical assessment.

Future work will implement a community validation module that allows direct assessment and annotation by community members. We will continue exploring expansion through new partnerships and investigate how structured knowledge can inform AI systems by enabling them to defer to community-approved sources with proper attribution. 

As part of this ongoing work, LLM-assisted identification components (e.g., LlamaExtract) may introduce hallucinated or misattributed relationships; preliminary qualitative inspection identified such cases, which require continued human oversight and will be further investigated in future research.


\begin{credits}
\subsubsection{\ackname}
This work was supported by funding from Genome Canada, Genome Alberta, and Genome British Columbia through the Canada Earth BioGenome Project (311CBG) and Genomics in Society for Arctic Biodiversity and Conservation. We gratefully acknowledge the partnership and knowledge sharing of the Kitikmeot Heritage Society, and extend our sincere thanks to the Elders in Residence and members of the Iqaluktuuttiaq community. We also thank Emily Angulalik for her support as translator and coordinator.
\end{credits}

\small
\setlength{\itemsep}{0pt}
\setlength{\parskip}{0pt}

\end{document}